\documentclass[12pt]{article}
\usepackage[utf8]{inputenc}
\usepackage[british]{babel}
\usepackage{cmap}
\usepackage{lmodern}

\usepackage{amssymb, amsmath, amsthm}
\usepackage[a4paper,top=25mm,bottom=25mm,left=25mm,right=25mm]{geometry}
\usepackage{ragged2e}

\usepackage{authblk} % for headings
\usepackage{pifont}
\usepackage{graphicx}
\usepackage[dvipsnames,svgnames,table]{xcolor}
\usepackage[figuresright]{rotating}
\usepackage{xtab} % tackle the long tables
\usepackage{longtable} % tackle the long tables
\usepackage{xltabular}
\usepackage{multirow}
\usepackage{footnote}
\usepackage[stable]{footmisc}
\usepackage{chngpage} % allows for temporary adjustment of side margins
\usepackage{pdflscape} % landscape environment
\usepackage[nottoc,notlot,notlof]{tocbibind} % includes everything in ToC

\usepackage{pgfplots}
\pgfplotsset{compat=1.14}
\pgfplotsset{every tick label/.append style={font=\footnotesize}}
\usepackage{setspace}
\makesavenoteenv{tabular}
\usepackage{tabularx}
\usepackage{ltablex}
\usepackage{booktabs}
\usepackage{threeparttable} % [flushleft]
\usepackage[referable]{threeparttablex} % footnotes in tabu
\newcolumntype{R}{>{\raggedleft\arraybackslash}X}
\newcolumntype{L}{>{\raggedright\arraybackslash}X}
\newcolumntype{C}{>{\centering\arraybackslash}X}
\newcolumntype{A}{>{\columncolor{gray!25}}C}
\newcolumntype{a}{>{\columncolor{gray!25}}c}

\newlength{\tablen}

\usepackage{dcolumn} % alignment to decimal points
\newcolumntype{.}{D{.}{.}{-1}}

\usepackage{tikz}
\usetikzlibrary{arrows, calc, matrix, patterns, positioning, shapes, trees}
\usepackage[semicolon]{natbib}
\usepackage[hyphens]{url}
\usepackage{hyperref} % [hidelinks]
\hypersetup{
  colorlinks   = true,    % Colours links instead of ugly boxes
  urlcolor     = blue,    % Colour for external hyperlinks
  linkcolor    = blue,    % Colour of internal links
  citecolor    = ForestGreen      % Colour of citations
}
\usepackage{microtype}
\usepackage[singlelinecheck=false,justification=centerfirst]{caption} % [justification=centering]
\usepackage[toc,page]{appendix}

\usepackage[labelformat=simple]{subcaption}

\DeclareCaptionLabelFormat{parenthesis}{(#2)}
\makeatletter
\renewcommand\p@subfigure{\arabic{figure}.}
\makeatother

\DeclareCaptionLabelFormat{parenthesis}{(#2)}
\makeatletter
\renewcommand\p@subtable{\arabic{table}.}
\makeatother

\usepackage{enumitem}
\setlist[itemize]{leftmargin=2.5\parindent}
\setlist[enumerate]{leftmargin=2.5\parindent}

\def\addlegendimage{\csname pgfplots@addlegendimage\endcsname}

\theoremstyle{plain}
\theoremstyle{definition}
\newtheorem{definition}{Definition}%[section]
\newtheorem{example}{Example}%[section]

\theoremstyle{remark}

\def\keywords{\vspace{.5em} % Add keywords
{\noindent \textit{Keywords}: }}

\def\JEL{\vspace{.5em} % Add keywords
{\noindent \textbf{\emph{JEL} classification number}: }}

\def\AMS{\vspace{.5em} % Add keywords
{\noindent \textbf{\emph{MSC} class}: }}

\title{Optimising the decision threshold in a weighted voting system: The case of the IMF's Board of Governors}
\author{
\href{https://sites.google.com/view/doragretapetroczy}{D\'ora Gr\'eta Petr\'oczy}\thanks{~E-mail: \emph{apetroczy@metropolitan.hu} 
\newline Magyar Nemzeti Bank, Budapest, Hungary
\newline Budapest Metropolitan University, Budapest, Hungary}
}

\date{\today}

\begin{document}

\maketitle
\begin{abstract}
\noindent
In a weighted majority voting game, the players' weights are determined based on the constitutional planner's intentions. The weights are challenging to change in numerous cases, as they represent some desired disparity. However, the voting weights and the actual voting power do not necessarily coincide. Changing a decision threshold would offer some remedy. 
The International Monetary Fund (IMF) is one of the most important international organisations that uses a weighted voting system to make decisions. The voting weights in its Board of Governors depend on the quotas of the 191 member countries, which reflect their economic strengths to some extent. We analyse the connection between the decision threshold and the a priori voting power of the countries by calculating the Banzhaf indices for each threshold between 50\% and 87\%. The difference between quotas and voting powers is minimised if the decision threshold is 58\% or 59\%.
% and determines their contribution to IMF expenditure and the amount of loans they can borrow from the organisation
\end{abstract}


\AMS{91A80, 91B12}
% Game Theory, Applications of game theory
% Voting theory

\JEL{C71, D72}
% Cooperative Games
% Analysis of Collective Decision-Making: Political Processes: Rent-Seeking, Lobbying, Elections, Legislatures, and Voting Behavior

\maketitle

\section{Introduction}
Even if the weights are allocated fairly, a country's real voting power does not necessarily coincide with its voting weight. The most famous case was probably Luxembourg in the Council of the European Union during the period 1958-1972: the country had a positive voting weight without any influence on the outcome of the voting \citep{BramsAffuso1976}.

Changing voting weights is difficult, as they are usually tied to underlying factors or agreements. In contrast, the decision-making threshold is generally not subject to such constraints.

In our research, we examine a specific case ---the IMF (International Monetary Fund) Board of Governors --- to analyse how the discrepancy between voting weights and voting power changes as a result of variations in the decision-making threshold. We also investigate how inequality among voters evolves and how these changes influence the Board’s ability to reach decisions.

The IMF, an international financial institution with 191 member countries, has two main decision-making bodies: the Board of Governors and the Executive Board. The Board of Governors decides on major issues but may delegate certain reserved rights to the Executive Board. Each member country appoints a governor to the Board of Governors, usually the finance minister or the central bank governor. 
%It Meetings are usually held annually.

In the Board of Governors, a weighted voting system is used for decision-making, usually with a 50\% threshold, but 85\% for some important issues.
An IMF member's quota determines its voting weight, calculated by basic votes (equal for all) plus one vote per Special Drawing Rights  (SDRs) 100,000 of quota. The SDR is an international reserve asset; it is not a currency, but its value is based on a basket of five currencies: the US dollar, the euro, the Chinese renminbi, the Japanese yen, and the British pound sterling.
The 16th quota revision took place in December 2023, with the first increase of voting weights in a long time. However, it is effective only from the end of 2024.

The naming is a bit misleading. In the IMF, a country’s `quota' determines its financial contribution and borrowing power, whereas in voting literature, `quota' usually refers to the threshold required to reach a decision. Also, the IMF calls voting power the share of voting weights. For clarity, in this article, we will refer to the original IMF voting power as weights and the decision quotas as thresholds.

Although the 14th General Review of Quotas represented a major step toward reflecting the increasing role of dynamic emerging markets and developing countries in the institution's governance structure, the actual voting weight shares still differ substantially from the so-called calculated quota shares determined by objective economic variables \citep{Tran2024}.

The current paper explores the relationship between voting weights and the majority threshold under the new rules.  While our calculations are similar to those of \citet{Leech2002b}, we utilise more recent data and more precise power index values. \citet{Kurz2016} calculates voting power using different thresholds but does not compare it to the weights. 
\citet{BhattacherjeeSarkar2019} apply a similar methodology but for the IMF's Executive Directors. To measure the difference between the weights and the power indices, we propose a new approach: the maximal ratio \citep{Csato2024}.

 The \emph{a priori} voting powers are quantified by the Banzhaf index, which is one of the most popular indicators of voting power: \citet{Leech2002b} used it to measure the distribution of voting power in the IMF, and several studies used it in the case of the Council of the European Union \citep{FelsenthalMachover1997, LaruelleWidgren1998, FelsenthalMachover2001, Leech2002a, NapelWidgren2011, Kirsch2022}.

Our central finding is that a decision threshold of 58\% or 59\%, depending on the distance measure, would minimise the difference between the voting weights and the voting powers of the 191 members.

Therefore, we have an important message for decision-makers: during future negotiations, it is not sufficient to focus only on the voting weights; the decision threshold should also receive attention. Otherwise, the seemingly fair allocation of voting weights might favour some countries at the expense of others in real voting situations. 

%The 14th General Review of Quotas was part of reforms to IMF quotas and governance. It was completed in 2010 and became effective in 2016. The reforms represented a major step toward better reflecting the increasing role of dynamic emerging markets and developing countries in the institution's governance structure.

%The 15th General Review of Quotas concluded in 2020 with no quota increase and guided the 16th Review. The Board of Governors concluded the 16th quota review in December 2023, approving an increase in quotas by 50 percent. The next step is for member countries to consent to their respective quota increases. However, increasing the quotas with the same ratio does not change the power ratios in voting. The IMFC Chair’s Statement also calls for
%“work to develop, by June 2025, possible approaches as a guide for further quota realignment,
%including through a new quota formula, under the 17th General Review of Quotas”\footnote{~\url{https://www.imf.org/en/News/Articles/2023/11/07/pr23383-imf-executive-board-approves-a-proposal-to-increase-imf-quotas}}.
%Thus it is worth it to investigate the real power in the voting.

%Using the Banzhaf power index we measure the ex-ante voting power of countries in the IMF. The Banzhaf index is one of the most common indicators, \citet{Leech2002b} used to measure the distribution of voting power in the IMF, \citet{Leech2002a, Gollner2017, Szczypinska2018, Gabor2020, Kirsch2022} calculate it for the case of Council in the European Union.

%We provide a sensitivity analysis regarding the quota. We have found that increasing the quota from 50\% to 59\% would decrease significantly the distance between the voting weights and ex-ante voting power.

Beyond its empirical focus, the paper makes a conceptual contribution to the analysis of weighted voting games by reframing the decision threshold as an endogenous design parameter rather than as an exogenously fixed institutional constant. While existing studies typically evaluate voting power under given quotas, our approach examines how to optimally choose quotas to align voting power with normatively intended voting weights. In this sense, the paper contributes to the design-oriented literature on voting rules by proposing quota-power consistency as an operational fairness objective in large, highly asymmetric weighted voting systems.

The paper is structured as follows.
Section~\ref{Sec2} provides a concise overview of related papers.
Section~\ref{Sec3} introduces the basic concepts of our approach, which are used to carry out the sensitivity analysis in Section~\ref{Sec4}. Finally, Section~\ref{Sec5} ends with concluding remarks.

\section{Literature review}\label{Sec2}
There are many fairness concepts in the literature.
\citet{BeisbartBovens2007} provide a foundational distinction between utilitarian fairness--- maximising expected total utility--- and egalitarian fairness, which aims to equalise individuals’ expected utilities. They show that if the interests of people in the same constituency are uncorrelated, then a weighted rule with square-root weights does best with respect to both ideals. If
there are perfect correlations, then the utilitarian ideal requires proportional weights,
whereas the egalitarian ideal requires equal weights. 

This welfare-based notion of fairness is refined by \citet{Koriyamaetal2013}, who offer a general utilitarian justification for degressive proportionality. 

Subsequent work shifts the fairness criterion from welfare outcomes to democratic representation and individual influence. \citet{MaaserNapel2012} evaluating voting weights by how closely representative decisions approximate those of direct democracy. They show that the fairness-optimal weighting rule depends on preference heterogeneity: square‑root weights are fair under i.i.d. preferences, while proportional weights become fairer when voters are more similar within constituencies. 

This influence-oriented interpretation is formalised and generalised in \citet{KurzMaaserNapel2017}, who explicitly ground fairness in the democratic principle of one person, one vote. Defining fairness as equality of a priori individual influence, they analyse pivot probabilities in a two-tier median voter model over continuous policy spaces and show that fair representation requires square-root weights only under independence, while realistic polarisation across constituencies robustly restores proportional weighting.

The weighted voting system of the IMF has attracted considerable attention in the academic literature.

\citet{FischerSchotter1978} proved that no weighted voting system is free from the paradox of redistribution, that is, a lower voting weight may increase voting power if there are at least seven voters. \citet{DreyerSchotter1980} examined the effect of the Second Amendment to the Articles of Agreement of the IMF. Under the proposed change, the voting weight of 38 countries decreased, yet their voting power increased when measured by the Banzhaf power index.

\citet{Leech2002b} provides evidence that the power of the United States is much greater than its nominal share of the votes, and the special 85\% majority requirement for major decisions severely limits the effectiveness of the decision-making system. The United States is also most affected if the majority requirement changes, as its power index declines steeply when the threshold rises. \citet{Leech2003} provide a new method for approximate power indices in large voting games, which makes it possible to analyse the voting power in the IMF.

\citet{StrandRapkin2005} examine the potential influence of hypothetical regional voting blocs within the IMF. They assess using three power indices--- Banzhaf, Johnston, and Shapley--Shubik--- how regional coalitions like the EU, ASEAN+3, and APEC could shift voting power dynamics in the IMF’s Executive Board. Their analysis reveals that the United States maintains dominant voting power in nearly all scenarios, even though a unified European bloc could rival or surpass its influence. Conversely, Asian countries, even when grouped, struggle to match the power of United States unless they align with broader coalitions like APEC, which ironically would still be dominated by the United States. 

\citet{RapkinStrand2005} analyse how developing countries are represented in the IMF’s quota and voting system. While these countries often appear over-represented using the IMF’s traditional economic indicators, their actual influence is found to remain limited due to low voting shares and structural imbalances. \citet{RapkinStrand2006} assess various reform proposals and advocate for a double majority voting system (requiring both a majority of member states and a majority of weighted votes) to better balance sovereign equality with economic power, similarly to the Council of the European Union. 

\citet{AlonsoMeijideBowles2005} use coalitional power indices for the Executive Directors. Efficient computational methods using generating functions are introduced to calculate coalitional power indices in weighted majority games with \emph{a priori} unions, such as those found in the IMF’s Executive Board.  Their study reveals significant differences in power measurement when coalition structures are considered.

\citet{LeechLeech2006} analyse the governance structures of the IMF and World Bank using power indices. The weighted voting systems in these institutions are shown to disproportionately amplify the power of the United States beyond its voting weight, while diminishing the influence of other members, especially developing countries. The authors also explore the effects of restoring ``basic votes” to their original 1946 level and find that while this would improve representation for poorer countries, the United States would still retain outsized power. Additionally, the paper criticises the constituency system for granting excessive influence to small European countries such as Belgium and the Netherlands.

\citet{AleskerovKalyaginPogorelskiy2008}  present a novel approach to measuring the voting power of IMF member countries by incorporating their preferences to form coalitions based on political-economic integration and regional proximity. Unlike traditional indices such as Banzhaf or Penrose, the authors introduce two new absolute power indices that account for the two-tiered structure of IMF decision-making, both within constituencies and at the Executive Board level. They calculate a compound power index as the product of a country’s influence within its constituency and its constituency’s influence in the Executive Board. Actual voting power can differ significantly from nominal voting weights, especially under higher decision thresholds (e.g., 85\%). Notably, under such thresholds, the Netherlands’ constituency surpasses the United States in power, contrary to what classical indices suggest. Their indices are robust to variations in preference weights. \citet{AleskerovKalyaginPogorelskiy2010}  extend these results by introducing a new preference-based model that incorporates bilateral trade data to estimate countries’ willingness to form coalitions. They investigate how these preferences affect actual voting power within the IMF’s Executive Board under different majority thresholds (simple, 70\%, and 85\%). Their findings show that as the decision threshold increases, the influence of coalition preferences becomes more significant, especially for smaller countries.

\citet{LeechLeech2013} use voting power analysis for both the Board of Governors and the Executive Board to show that the power of the United States is enhanced at the expense of all other members.

\citet{FreixasKaniovski2014} introduce the Minimum Sum Representation (MSR) index as a novel measure of voting power in weighted voting games and apply it to analyse two IMF constituencies. The MSR index distributes power more equally than traditional indices, assigning less power to dominant countries (e.g., Belgium) and more to smaller ones (e.g., Kosovo and Slovenia), resulting in a lower Gini coefficient and a fairer power distribution.

\citet{Clark2017} investigates the unequal distribution of voting weights within the IMF and the World Bank from 1951 to 2010. Despite global economic convergence, voting weights in these institutions remains highly unequal, marginalising developing countries, especially those with high debt burdens. The research finds that beyond official economic indicators like GDP, social networks, military alliances, and world-system positions also significantly influence voting weight allocation. 

\citet{BhattacherjeeSarkar2019} propose using Pearson’s correlation coefficient to assess the similarity between weights and power indices, and standard inequality indices (like the Gini index and coefficient of variation) to compare inequality in weights versus power. Since the choice of voting threshold significantly affects both similarity and inequality, they argue that adjusting this threshold, rather than altering the weights, can better align voting power with intended socio-economic representations. Applying their framework to the IMF Executive Board, they find the current threshold to be suboptimal.

\citet{MayerNapel2020} analyse how voting procedures affect the expected success of Executive Directors in the IMF when selecting the Managing Director from a shortlist of three candidates. Using \emph{a priori} assumptions and Monte Carlo simulations, they compare three voting rules—plurality, plurality with runoff, and Copeland (pairwise majority)—to assess how closely outcomes align with the preferences of each Director. The United States, with the largest voting share, benefits most from simple plurality, while other Directors fare better under runoff or Copeland rules. 

\citet{KurzMayerNapel2021}  extend classical voting power analysis to multi-alternative decisions in weighted committees, introducing generalised versions of the Penrose--Banzhaf and Shapley--Shubik indices. These new indices measure a player's influence as the probability that a random change in their preferences alters the collective outcome, accounting for various voting rules such as plurality, Borda, Copeland, and runoff systems. They apply their framework to real-world cases, including the IMF Executive Board's election of the Managing Director, demonstrating that the choice of voting rule significantly affects the distribution of influence, often more than changes in the voting weights.

Our research is the first to compare the voting powers with the weights in the case of different quotas, while also considering the decisiveness  of the suggested voting threshold. \citet{BhattacherjeeSarkar2019} conduct similar research, but only for the Executive Directors, using different distance measures.

Our findings can be interpreted as numerical and institutional reinforcement of the analytical optimal-quota result of \citet{Ruff2009}. Despite the IMF’s highly unequal and non-idealised weight structure, the empirically optimal threshold emerges within a remarkably similar range.

\section{Methodology and data}\label{Sec3}

Voting situations are usually modelled by a cooperative game with transferable utility where the voters are the players, and the value of a coalition is maximal if it can accept a proposal and minimal otherwise.

Let $N$ denote the set of players and $S \subseteq N$ be a coalition. The cardinality of a set is denoted by the corresponding lowercase letter; for example, the number of players in coalition $S$ is $\lvert S \rvert = s$, and the number of players is $\lvert N \rvert = n$. The value of any coalition is given by the characteristic function $v: 2^N \to \mathbb{R}$.

\begin{definition}
\emph{Simple voting game}:
A game $(N,v)$ is called a simple voting game if
\[
v(S) \in \left\{ 0,1 \right\}  \text{ for all $S \subseteq N$}.
\]
\end{definition}

Let vector $\mathbf{w} \in \mathbb{R}_+^n$ denote the weights of the players and $q \in \mathbb{R_+}$ denote the decision threshold.

\begin{definition}
\emph{Weighted voting game}:
A game $(N,v)$ can be represented as a weighted voting game $[\mathbf{w},q]$ if and only if for any coalition $S \subseteq N$:

\[
v(S)= \left\{ 
\begin{array}{cl}
1& \text{if }   \sum_{j \in S} w_j \geq q \\ 
0& \text{otherwise}.
\end{array}
\right.
\]
\end{definition}
Coalitions $S$ such that $v(S)=1$
 are called winning coalitions, while coalitions $S$ as $v(S)=0$  are the losing ones.
A player is called \textit{critical} if it can turn a winning coalition into a losing one. The  Banzhaf index \citep{Banzhaf1964, Coleman1971, Penrose1946}, shows the probability that a player influences a decision by taking into account the ratio of coalitions where the player is critical. As we have already seen, the Banzhaf index is one of the most common measures of \emph{a priori} voting power.

\begin{definition} \emph{Absolute Banzhaf index}:
Let $(N,v)$ be a simple voting game.
The absolute Banzhaf index of player $i \in N$ is:
\[
\sum_{S  \subseteq N \setminus  \{i\}}  \frac{1}{2^{n-1} }\left(v\big(S\cup\left\{i\right\}\big)-v\big(S\big)\right)= \frac{\eta_i (N, v)}{2^{n-1}},
\]
where $\eta_i(v)$ is the Banzhaf score of player $i$, the number of coalitions where $i$ is critical.
\end{definition}

We use its normalised version as the measure of voting power.

\begin{definition}\emph{(Normalised) Banzhaf index}:
Let $(N,v)$ be a simple voting game.
The Banzhaf index of player $i \in N$ is its normalised Banzhaf score:
\[
\beta_i(N,v) =\frac{\eta_i (N,v)}{\sum_{j\in N} \eta_j (N,v)}.
\]
\end{definition}

Another widely discussed power index is the Shapley--Shubik index \citep{ShapleyShubik1954}.
\begin{definition}
\emph{Shapley--Shubik index}:
Let $(N,v)$ be a simple voting game.
The Shapley--Shubik index of player $i$ is
\[
\varphi_i(N,v) = \sum_{S  \subseteq N \setminus  \{i\}}  \frac{s!(n-s-1)!}{n!} \left[ v \big( S \cup \left\{ i \right\} \big) -v \left( S \right) \right].
\]
\end{definition}
Consider a random order of the players. The Shapley--Shubik index of a player is its marginal contribution to the coalition formed by the preceding players, averaged over the set of all the possible orders of the players.

In many situations, the two power indices shows similar patterns. However, in many cases, they disagree seriously. \citet{Straffin1977} provides a concise summary.  The disagreement persists in the case of near-oceanic games, games with several powerful players and a large number of symmetric weak players. 
\begin{example}\label{Examp1}
Consider a major voter, with 20\% of the weights, and the remainder is split equally among a large number of minor voters.

\begin{table}[t!]
  \centering
  \rowcolors{1}{}{gray!20} 
  \caption{The value of power indices in case of near-oceanic games}
  \label{Table1_oceanic}
\begin{threeparttable}[t!]
    \begin{tabularx}{0.9\textwidth}{l RRRR} \toprule \hiderowcolors
    No. minor voters& $\mathit{NBI_{major}}(\%)$& $\mathit{NBI_{minor}}(\%)$& $\mathit{SSI_{major}}(\%)$& $\mathit{SSI_{minor}}(\%)$\\    \bottomrule \showrowcolors
    50& 47.67& 1.04& 25.49& 1.49\\
    100& 74.40& 0.26& 24.75&0.75\\
    150& 91.11& 0.06& 24.50& 0.50\\
    200&97.74& 0.01& 24.88&  0.38\\
    250& 99.59& 1.65e-03& 25.10& 0.30\\
    300& 99.89& 3.54e-04& 24.92& 0.25\\
 \bottomrule  
\end{tabularx} 
\begin{tablenotes} \footnotesize	
\item
\emph{Note:} Major voter has 20\% of the weights, threshold=50\%
\end{tablenotes}
\end{threeparttable}
\end{table}    
\end{example}

If the number of minor voters tends to infinity, with 50\% decision threshold, the limit of the Shapley--Shubik index is $\frac{w}{1-w}$, where $w$ is the weight share of the major voter \citep{ShapiroShapley1978}. For the Banzhaf index, the limit is 1, as Table~\ref{Table1_oceanic} shows with different number of minor players.

%\citet{Leech2002c} shows in real-life situations, where there is a few large shareholder, and many minor one, the Shapley--Shubik index of the largest shareholder is rarely as high as independent evidence suggests it should be given the distribution of voting weights. Therefore using the Shapley--Shubik index as a measure in these games is not recommended. In our analysis, we limit ourselves to the Banzhaf index.
\citet{Leech2002c} shows that in real-life voting bodies with a small number of large shareholders and a long tail of minor ones, the Shapley–Shubik index tends to underestimate the actual influence of the largest member (such as observed voting behaviour, governance outcomes, or institutional practice).  The IMF Board of Governors itself is a clear example of such a structure, with a few major members holding substantial shares and many countries possessing very small voting weights. Therefore, using the Shapley--Shubik index as a measure in these games is not recommended. In our analysis, we limit ourselves to the Banzhaf index.

%To calculate these indices we use Dennis Leech's program \href{https://homepages.warwick.ac.uk/~ecaae/ipmmle.html}{ipmmle}.  The algorithm uses the modification of the approximation method of multilinear extensions of Guillermo Owen (see \citet{Owen2013}) described in \citet{Leech2003}.

\subsection{Calculation algorithm}
To calculate these indices, we use the \href{https://pypi.org/project/power-bdd/}{power-bdd python package},  which calculates the power indices by writing the game in a binary decision diagram (BDD) form \citep{Bolus2011}. The straightforward calculation of the Banzhaf index for all players has a time-complexity of order $n2^n$, while \citet{Bolus2011} algorithm has $O(nq \log(q))$ for one-tier games \citep{Wilms2020}.

%Unfortunately, the calculations are not feasible with the original weights, as they would require more memory than our current resources can provide. Therefore, we took each country’s percentage share, multiplied it by one hundred, and rounded it to the nearest integer. For example, the United States holds 16.49\% of the total weight (831394), so in the computations, we assign it a weight of 1649. However, in this transformed system, Belize, with a weight of 1719, and Tuvalu, with a weight of 1477, both end up receiving the value 3 in our calculations.
\subsection{Weight comparison}

In order to compare the normalised weights ($w$) and the Banzhaf indices ($\beta$) three metrics are considered:

\begin{itemize}
    \item Euclidean distance:
    
        $d_{euc}(\beta, w )=\sqrt{\sum_{i=1}^{n}(\beta_i-w_i)^2}$
    \item Manhattan distance:

    $d_{man}=\sum_{i=1}^n \left| \beta_i-w_i \right| $
   
    \item Maximal ratio \citep{Csato2024}:

    $\omega(\beta, w )=\max \biggl\{ \max \Bigl\{ \frac{\beta_i}{w_i}  ; \frac{w_i}{\beta_i} : 1\le i \le  n \Bigl\}\biggl\}$

\end{itemize}
Euclidean and Manhattan distances fail to capture substantial relative deviations in the weights and Banzhaf indices of members with small voting weights. In contrast, the maximal ratio criterion ensures that no country is disproportionately advantaged (high Banzhaf-to-weight ratio) or disadvantaged (low Banzhaf-to-weight ratio) in terms of voting power.

\begin{example}\label{examp1}
    Let $\mathbf{u} = [0.5,\ 0.4,\ 0.1]$  and $\mathbf{v} = [0.5,\ 0.3,\ 0.2]$ 
two vectors. In this case, $d_{\text{euc}}(\mathbf{u}, \mathbf{v}) = \sqrt{(0.5 - 0.5)^2 + (0.4 - 0.3)^2 + (0.1 - 0.2)^2} = \sqrt{0 + 0.01 + 0.01} = \sqrt{0.02} \approx 0.1414$; $d_{\text{man}}(\mathbf{u}, \mathbf{v}) = |0.5 - 0.5| + |0.4 - 0.3| + |0.1 - 0.2| = 0 + 0.1 + 0.1 = 0.2$ and maximal ratio is $\max \{1,1.333,2\}=2$.

\end{example}
Example~\ref{examp1} shows that while the Euclidean and the Manhattan distance handle the difference between $(0.4, 0.3)$ and $(0.1, 0.2)$ as the same, the maximal ratio focuses on the relative difference.

\citet{BhattacherjeeSarkar2019} use Pearson's correlation to capture the closeness between the weights and the value of the Banzhaf power indices. 
We also take into account this approach. 

\begin{definition}\emph{Pearson's correlation}: The correlation between weights ($w$) and power indices $\beta$ are:
\[
\rho(w,\beta)=\frac{cov(w,\beta)}{\sigma_w \sigma_{\beta}}
\]
where
\begin{itemize}
\item cov is the covariance
\item $\sigma_w$  is the standard deviation of $w$
\item $\sigma_\beta$  is the standard deviation of $\beta$
\end{itemize}

\end{definition}

It is also worth examining how the probability of decision-making changes with the quota, meaning what percentage of all possible coalitions can decide on a proposal. We use the Power of the Body to Act (PTA) index to measure decisiveness \citep{Coleman1971}.
\begin{definition}
\emph{Power of the Body to Act}:
Let $\left( N,v \right)$ be a simple voting game.
The Power of the Body to Act index is the number of winning coalitions divided by the number of possible
coalitions $2^n$.

\end{definition}
For example, a unanimity requirement would give a value of PTA
equal to $1/2^{n}$, while a quota of  $50\%+\epsilon$ would give PTA around $0.5$.
 This Power of the Body to Act has been axiomatically characterised by \citet{BaruaChakravartyRoy2009}.

%Another possibility to find the optimal decision threshold is to focus on the level of inequality, such that the baseline is the inequality implied by the voting weights. The most widely used inequality indicators are the Herfindahl-Hirschmann and the Gini indices.
%\begin{definition}\emph{Gini index}:
%The value of the Gini index of a vector $\mathbf{a} = (a_1, ... ,a_n)$  is given by 
%\[
%Gini(\mathbf{a})=\frac{\sum_{i=1}^{n}\sum_{j=1}^{n}\lvert a_i-a_j \rvert}{2n\sum_{i=1}^{n}a_i}.
%\]  
%\end{definition}

%\begin{definition}\emph{Herfindahl-Hirschmann index}: The value of the Herfindahl-Hirschmann index of a vector $\mathbf{a} = (a_1, ... ,a_n)$  is given by 
%\[
%HHI(\mathbf{a}) = \sum_{i=1}^n a_i^2.
%\]    
%\end{definition}

\subsection{Decision making in the IMF}

Each member country is assigned a quota, which determines its participation in the IMF's capital and voting power.
The quotas serve four roles in the IMF's operations, and are denominated in SDRs, the IMF’s unit of account.  Quotas determine the maximum amount of financial resources a member is obliged to provide. They are a key determinant of voting power in IMF decisions. Members get one vote per SDR 100,000 of quota, plus basic votes, which are the same for all members. They determine the maximum amount of loans a member can obtain under normal access, and determine a member’s share in a general allocation of SDRs.

A quota formula is used to help assess members’ relative position in the world economy, and it can play a role in guiding the distribution of quota increases. 
The current formula is a weighted average of GDP (weight of 50 percent), openness to the global economy (30 percent), economic variability (15 percent), and international reserves (5 percent) \citep{IMF2022}.

The voting weights for our calculations are collected from the \href{https://www.imf.org/en/About/executive-board/members-quotas}{IMF} website and provided in Table~\ref{Table_A1} in the Appendix.

\section{Results}\label{Sec4}

\begin{table}[t!]
  \centering
  \rowcolors{1}{}{gray!20} 
  \caption{The voting weights and powers of the top 10 countries with a decision threshold of 50\% and 85\%}
  \label{Table1}
\begin{threeparttable}[t!]
    \begin{tabularx}{0.8\textwidth}{l RRR} \toprule \hiderowcolors
    Country& Weight& $\mathit{NBI}$(50\%)& $\mathit{NBI}$(85\%) \\    \bottomrule \showrowcolors
    United States& 16.49\%& 23.26\%& 2.90\% \\
    Japan& 6.14\%& 5.54\%& 2.90\% \\
    China& 6.08\%& 5.49\%& 2.90\% \\
    Germany& 5.31\%& 4.84\%& 2.90\% \\
    France& 4.03\%& 3.70\%& 2.89\% \\
    United Kingdom& 4.03\%& 3.70\%& 2.89\% \\
    Italy& 3.02\%& 2.78\%& 2.81\% \\
    India& 2.63\%& 2.43\%& 2.73\% \\
    Russia& 2.59\%& 2.39\%& 2.72\% \\
    Brazil& 2.22\%& 2.05\%& 2.58\% \\ \bottomrule  
\end{tabularx} 
\begin{tablenotes} \footnotesize	
\item
\emph{Note:} $\mathit{NBI}$(50\%) denotes the value of Banzhaf index, with 50\% threshold
\end{tablenotes}
\end{threeparttable}
\end{table}

Table~\ref{Table1} shows the power of the top 10 countries with the largest voting weight. Under the current rule with a 50\% quota, the power of the United States is far higher than its weight, while the other major countries have less influence.

\input{Fig1_Banzhaf_indices_top10}

Figure~\ref{Fig1} shows how these countries' power changes with the quota. In the case of the United States, the power-threshold relation decreases steeply for quotas above 50\%. For the other countries, there is an M-shape pattern.

The United States exhibits a single‑peaked Banzhaf–threshold curve because its voting weight is so large relative to all other members. At low thresholds (around 50\%), the U.S. is critical in an extremely high proportion of winning coalitions, which gives it disproportionately high voting power. As the threshold increases, more coalitions either no longer need the U.S. to reach the threshold or become so large that removing the U.S. does not change their winning status. Consequently, the number of coalitions in which the U.S. is critical falls monotonically, producing a smooth, unimodal curve without secondary peaks.
By contrast, medium‑sized countries have voting weights in the 2–6\% range, as the threshold changes, these countries alternate between being critical and non‑critical depending on how coalition structures adjust. At moderate thresholds, they may lose influence because larger coalitions have multiple substitutes for a single medium‑sized player. However, at higher thresholds, large countries alone cannot meet them, and small countries contribute too little, forcing coalitions to include several medium‑sized members, making them critical again. This dynamic creates two separate regions of high criticality, producing the characteristic M‑shaped pattern in their Banzhaf indices.
Figure~\ref{Fig5} shows the power graphs of three medium-small countries,  Norway (voting weight 0.77\%; ranked 26th, Luxembourg (0.29\%, 55th),  and Iceland (0.09\%, 100th).  Norway and Luxembourg are on the same M-shaped curve, as in the case of medium-sized countries, but the ``hump" is noticeably farther from the middle. Iceland has only one peak with a high threshold, meaning, close to unanimity, its power is larger than the equal.
Figure~\ref{Fig6} shows three small countries, Malawi (0.06\%, 145th), Djibouti (0.04\%, 165th) and Timor-Leste (0.03\%, 168th). Their Banzhaf indices are an order of magnitude smaller than those in Figure~\ref{Fig5} and remain close to zero over much of the threshold interval. This is consistent with the structural disadvantage of small members in weighted voting games: because their weights are tiny relative to the threshold, they rarely constitute the decisive increment that transforms a losing coalition into a winning one.  As the threshold increases, the set of winning coalitions shifts toward larger, more inclusive groupings, and small members may become critical slightly more often, particularly when coalitions are just at the margin of the threshold, and additional small contributions matter.

\begin{figure}[t!]
\centering

\begin{tikzpicture}
\begin{axis}[
title style = {font=\small},
xlabel = Threshold (\%),
x label style = {font=\small},
ylabel = Banzhaf index (\%),
y label style = {font=\small},
%y tick label style = {/pgf/number format/.cd,fixed,fixed zerofill,precision=2},
width = \textwidth,
height = 0.6\textwidth,
ymajorgrids = true,
xmin = 1,
xmax = 100,
ymin = 0,
%ymax = 4.1,
%max space between ticks=50,
legend style = {font=\small,at={(0.25,-0.15)},anchor=north west,legend columns=5},
legend entries = {Norway$\qquad$,Luxembourg$\qquad$,Iceland}
] 
% Norway
\addplot [red, mark=pentagon, mark size=2pt, mark options=solid] coordinates {
(1,0.6525)
(2,0.7851)
(3,0.9275)
(4,1.0691)
(5,1.1939)
(6,1.2903)
(7,1.3545)
(8,1.3884)
(9,1.3967)
(10,1.3852)
(11,1.36)
(12,1.3268)
(13,1.2899)
(14,1.2522)
(15,1.2153)
(16,1.1802)
(17,1.1469)
(18,1.1158)
(19,1.0868)
(20,1.0599)
(21,1.035)
(22,1.012)
(23,0.9909)
(24,0.9715)
(25,0.9536)
(26,0.9372)
(27,0.922)
(28,0.9079)
(29,0.8948)
(30,0.8824)
(31,0.8707)
(32,0.8596)
(33,0.8489)
(34,0.8385)
(35,0.8284)
(36,0.8186)
(37,0.8089)
(38,0.7993)
(39,0.7897)
(40,0.7803)
(41,0.7709)
(42,0.7616)
(43,0.7526)
(44,0.744)
(45,0.736)
(46,0.7289)
(47,0.7229)
(48,0.7184)
(49,0.7156)
(50,0.7146)
(51,0.7156)
(52,0.7184)
(53,0.7229)
(54,0.7289)
(55,0.736)
(56,0.744)
(57,0.7526)
(58,0.7616)
(59,0.7709)
(60,0.7803)
(61,0.7897)
(62,0.7993)
(63,0.8089)
(64,0.8186)
(65,0.8284)
(66,0.8385)
(67,0.8489)
(68,0.8596)
(69,0.8707)
(70,0.8824)
(71,0.8948)
(72,0.9079)
(73,0.922)
(74,0.9372)
(75,0.9536)
(76,0.9715)
(77,0.9909)
(78,1.012)
(79,1.035)
(80,1.0599)
(81,1.0868)
(82,1.1158)
(83,1.1469)
(84,1.1802)
(85,1.2153)
(86,1.2522)
(87,1.2899)
(88,1.3268)
(89,1.36)
(90,1.3852)
(91,1.3967)
(92,1.3884)
(93,1.3545)
(94,1.2903)
(95,1.1939)
(96,1.0691)
(97,0.9275)
(98,0.7851)
(99,0.6525)
(100,0.5236)

};
% Luxembourg
\addplot [ForestGreen, mark=star, mark size=2.5pt, mark options=solid] coordinates {
(1,0.6523)
(2,0.7737)
(3,0.852)
(4,0.8593)
(5,0.8236)
(6,0.7753)
(7,0.727)
(8,0.682)
(9,0.6413)
(10,0.6047)
(11,0.572)
(12,0.5431)
(13,0.5176)
(14,0.4952)
(15,0.4754)
(16,0.4578)
(17,0.4421)
(18,0.428)
(19,0.4152)
(20,0.4037)
(21,0.3933)
(22,0.3838)
(23,0.3752)
(24,0.3674)
(25,0.3603)
(26,0.3538)
(27,0.3478)
(28,0.3423)
(29,0.3371)
(30,0.3323)
(31,0.3278)
(32,0.3235)
(33,0.3194)
(34,0.3154)
(35,0.3116)
(36,0.3078)
(37,0.3042)
(38,0.3005)
(39,0.2969)
(40,0.2934)
(41,0.2898)
(42,0.2864)
(43,0.283)
(44,0.2798)
(45,0.2768)
(46,0.2742)
(47,0.2719)
(48,0.2703)
(49,0.2692)
(50,0.2688)
(51,0.2692)
(52,0.2703)
(53,0.2719)
(54,0.2742)
(55,0.2768)
(56,0.2798)
(57,0.283)
(58,0.2864)
(59,0.2898)
(60,0.2934)
(61,0.2969)
(62,0.3005)
(63,0.3042)
(64,0.3078)
(65,0.3116)
(66,0.3154)
(67,0.3194)
(68,0.3235)
(69,0.3278)
(70,0.3323)
(71,0.3371)
(72,0.3423)
(73,0.3478)
(74,0.3538)
(75,0.3603)
(76,0.3674)
(77,0.3752)
(78,0.3838)
(79,0.3933)
(80,0.4037)
(81,0.4152)
(82,0.428)
(83,0.4421)
(84,0.4578)
(85,0.4754)
(86,0.4952)
(87,0.5176)
(88,0.5431)
(89,0.572)
(90,0.6047)
(91,0.6413)
(92,0.682)
(93,0.727)
(94,0.7753)
(95,0.8236)
(96,0.8593)
(97,0.852)
(98,0.7737)
(99,0.6523)
(100,0.5236)

};
% Iceland
\addplot [black, mark=triangle, mark size=2pt, mark options=solid] coordinates {
(1,0.5619)
(2,0.4935)
(3,0.4142)
(4,0.3509)
(5,0.3054)
(6,0.2724)
(7,0.2473)
(8,0.2274)
(9,0.2109)
(10,0.1971)
(11,0.1853)
(12,0.1751)
(13,0.1664)
(14,0.1588)
(15,0.1522)
(16,0.1464)
(17,0.1412)
(18,0.1366)
(19,0.1325)
(20,0.1288)
(21,0.1254)
(22,0.1223)
(23,0.1196)
(24,0.117)
(25,0.1148)
(26,0.1127)
(27,0.1108)
(28,0.109)
(29,0.1074)
(30,0.1058)
(31,0.1044)
(32,0.103)
(33,0.1017)
(34,0.1004)
(35,0.0992)
(36,0.098)
(37,0.0968)
(38,0.0957)
(39,0.0945)
(40,0.0934)
(41,0.0923)
(42,0.0912)
(43,0.0901)
(44,0.0891)
(45,0.0881)
(46,0.0873)
(47,0.0866)
(48,0.086)
(49,0.0857)
(50,0.0856)
(51,0.0857)
(52,0.086)
(53,0.0866)
(54,0.0873)
(55,0.0881)
(56,0.0891)
(57,0.0901)
(58,0.0912)
(59,0.0923)
(60,0.0934)
(61,0.0945)
(62,0.0957)
(63,0.0968)
(64,0.098)
(65,0.0992)
(66,0.1004)
(67,0.1017)
(68,0.103)
(69,0.1044)
(70,0.1058)
(71,0.1074)
(72,0.109)
(73,0.1108)
(74,0.1127)
(75,0.1148)
(76,0.117)
(77,0.1196)
(78,0.1223)
(79,0.1254)
(80,0.1288)
(81,0.1325)
(82,0.1366)
(83,0.1412)
(84,0.1464)
(85,0.1522)
(86,0.1588)
(87,0.1664)
(88,0.1751)
(89,0.1853)
(90,0.1971)
(91,0.2109)
(92,0.2274)
(93,0.2473)
(94,0.2724)
(95,0.3054)
(96,0.3509)
(97,0.4142)
(98,0.4935)
(99,0.5619)
(100,0.5236)

};

% Current voting power
%\draw [black,very thick,dotted] (\pgfkeysvalueof{/pgfplots/xmin},1.368) -- (\pgfkeysvalueof{/pgfplots/xmax},1.368);
% Current population share
%\draw [brown,very thick,loosely dashed] (\pgfkeysvalueof{/pgfplots/xmin},0.863020998648138) -- (\pgfkeysvalueof{/pgfplots/xmax},0.863020998648138);
\end{axis}
\end{tikzpicture}

%\captionsetup{justification=centerfirst}
\caption{The voting power of Iceland, Luxembourg and Norway\\ }

\label{Fig5}

\end{figure}
\begin{figure}[t!]
\centering

\begin{tikzpicture}
\begin{axis}[
title style = {font=\small},
xlabel = Threshold (\%),
x label style = {font=\small},
ylabel = Banzhaf index (\%),
y label style = {font=\small},
%y tick label style = {/pgf/number format/.cd,fixed,fixed zerofill,precision=2},
width = \textwidth,
height = 0.6\textwidth,
ymajorgrids = true,
xmin = 1,
xmax = 100,
ymin = 0,
ymax = 0.43,
%max space between ticks=50,
legend style = {font=\small,at={(0.25,-0.15)},anchor=north west,legend columns=5},
legend entries = {Malawi$\qquad$,Djibouti 
$\qquad$,Timor-Leste
}
] 
% Malawi
\addplot [Orange, mark=pentagon, mark size=2pt, mark options=solid] coordinates {
(1,0.4249)
(2,0.3285)
(3,0.2623)
(4,0.2178)
(5,0.188)
(6,0.1669)
(7,0.1512)
(8,0.1387)
(9,0.1286)
(10,0.12)
(11,0.1128)
(12,0.1066)
(13,0.1013)
(14,0.0966)
(15,0.0926)
(16,0.0891)
(17,0.0859)
(18,0.0831)
(19,0.0806)
(20,0.0783)
(21,0.0763)
(22,0.0744)
(23,0.0727)
(24,0.0712)
(25,0.0698)
(26,0.0685)
(27,0.0674)
(28,0.0663)
(29,0.0653)
(30,0.0644)
(31,0.0635)
(32,0.0626)
(33,0.0618)
(34,0.0611)
(35,0.0603)
(36,0.0596)
(37,0.0589)
(38,0.0582)
(39,0.0575)
(40,0.0568)
(41,0.0561)
(42,0.0554)
(43,0.0548)
(44,0.0542)
(45,0.0536)
(46,0.0531)
(47,0.0526)
(48,0.0523)
(49,0.0521)
(50,0.052)
(51,0.0521)
(52,0.0523)
(53,0.0526)
(54,0.0531)
(55,0.0536)
(56,0.0542)
(57,0.0548)
(58,0.0554)
(59,0.0561)
(60,0.0568)
(61,0.0575)
(62,0.0582)
(63,0.0589)
(64,0.0596)
(65,0.0603)
(66,0.0611)
(67,0.0618)
(68,0.0626)
(69,0.0635)
(70,0.0644)
(71,0.0653)
(72,0.0663)
(73,0.0674)
(74,0.0685)
(75,0.0698)
(76,0.0712)
(77,0.0727)
(78,0.0744)
(79,0.0763)
(80,0.0783)
(81,0.0806)
(82,0.0831)
(83,0.0859)
(84,0.0891)
(85,0.0926)
(86,0.0966)
(87,0.1013)
(88,0.1066)
(89,0.1128)
(90,0.12)
(91,0.1286)
(92,0.1387)
(93,0.1512)
(94,0.1669)
(95,0.188)
(96,0.2178)
(97,0.2623)
(98,0.3285)
(99,0.4249)
(100,0.5236)

};
% Djibouti 
\addplot [ForestGreen, mark=star, mark size=2.5pt, mark options=solid] coordinates {
(1,0.2918)
(2,0.212)
(3,0.166)
(4,0.1368)
(5,0.1176)
(6,0.1043)
(7,0.0944)
(8,0.0866)
(9,0.0802)
(10,0.0749)
(11,0.0703)
(12,0.0665)
(13,0.0631)
(14,0.0602)
(15,0.0577)
(16,0.0555)
(17,0.0536)
(18,0.0518)
(19,0.0502)
(20,0.0488)
(21,0.0475)
(22,0.0464)
(23,0.0453)
(24,0.0444)
(25,0.0435)
(26,0.0427)
(27,0.042)
(28,0.0413)
(29,0.0407)
(30,0.0401)
(31,0.0396)
(32,0.039)
(33,0.0385)
(34,0.0381)
(35,0.0376)
(36,0.0371)
(37,0.0367)
(38,0.0363)
(39,0.0358)
(40,0.0354)
(41,0.035)
(42,0.0345)
(43,0.0341)
(44,0.0338)
(45,0.0334)
(46,0.0331)
(47,0.0328)
(48,0.0326)
(49,0.0325)
(50,0.0324)
(51,0.0325)
(52,0.0326)
(53,0.0328)
(54,0.0331)
(55,0.0334)
(56,0.0338)
(57,0.0341)
(58,0.0345)
(59,0.035)
(60,0.0354)
(61,0.0358)
(62,0.0363)
(63,0.0367)
(64,0.0371)
(65,0.0376)
(66,0.0381)
(67,0.0385)
(68,0.039)
(69,0.0396)
(70,0.0401)
(71,0.0407)
(72,0.0413)
(73,0.042)
(74,0.0427)
(75,0.0435)
(76,0.0444)
(77,0.0453)
(78,0.0464)
(79,0.0475)
(80,0.0488)
(81,0.0502)
(82,0.0518)
(83,0.0536)
(84,0.0555)
(85,0.0577)
(86,0.0602)
(87,0.0631)
(88,0.0665)
(89,0.0703)
(90,0.0749)
(91,0.0802)
(92,0.0866)
(93,0.0944)
(94,0.1043)
(95,0.1176)
(96,0.1368)
(97,0.166)
(98,0.212)
(99,0.2918)
(100,0.5236)

};
% Timor-Leste

\addplot [Periwinkle, mark=triangle, mark size=2pt, mark options=solid] coordinates {
(1,0.2828)
(2,0.2049)
(3,0.1602)
(4,0.132)
(5,0.1135)
(6,0.1007)
(7,0.0911)
(8,0.0835)
(9,0.0774)
(10,0.0722)
(11,0.0679)
(12,0.0641)
(13,0.0609)
(14,0.0581)
(15,0.0557)
(16,0.0536)
(17,0.0517)
(18,0.05)
(19,0.0485)
(20,0.0471)
(21,0.0459)
(22,0.0447)
(23,0.0437)
(24,0.0428)
(25,0.042)
(26,0.0412)
(27,0.0405)
(28,0.0399)
(29,0.0393)
(30,0.0387)
(31,0.0382)
(32,0.0377)
(33,0.0372)
(34,0.0367)
(35,0.0363)
(36,0.0358)
(37,0.0354)
(38,0.035)
(39,0.0346)
(40,0.0342)
(41,0.0337)
(42,0.0333)
(43,0.0329)
(44,0.0326)
(45,0.0322)
(46,0.0319)
(47,0.0317)
(48,0.0315)
(49,0.0313)
(50,0.0313)
(51,0.0313)
(52,0.0315)
(53,0.0317)
(54,0.0319)
(55,0.0322)
(56,0.0326)
(57,0.0329)
(58,0.0333)
(59,0.0337)
(60,0.0342)
(61,0.0346)
(62,0.035)
(63,0.0354)
(64,0.0358)
(65,0.0363)
(66,0.0367)
(67,0.0372)
(68,0.0377)
(69,0.0382)
(70,0.0387)
(71,0.0393)
(72,0.0399)
(73,0.0405)
(74,0.0412)
(75,0.042)
(76,0.0428)
(77,0.0437)
(78,0.0447)
(79,0.0459)
(80,0.0471)
(81,0.0485)
(82,0.05)
(83,0.0517)
(84,0.0536)
(85,0.0557)
(86,0.0581)
(87,0.0609)
(88,0.0641)
(89,0.0679)
(90,0.0722)
(91,0.0774)
(92,0.0835)
(93,0.0911)
(94,0.1007)
(95,0.1135)
(96,0.132)
(97,0.1602)
(98,0.2049)
(99,0.2828)
(100,0.5236)
};

% Current voting power
%\draw [black,very thick,dotted] (\pgfkeysvalueof{/pgfplots/xmin},1.368) -- (\pgfkeysvalueof{/pgfplots/xmax},1.368);
% Current population share
%\draw [brown,very thick,loosely dashed] (\pgfkeysvalueof{/pgfplots/xmin},0.863020998648138) -- (\pgfkeysvalueof{/pgfplots/xmax},0.863020998648138);
\end{axis}
\end{tikzpicture}

%\captionsetup{justification=centerfirst}
\caption{The voting power of Malawi,	Djibouti and Timor-Leste\\ }

\label{Fig6}

\end{figure}
\begin{figure}[t!]
\centering

\begin{tikzpicture}
\begin{axis}[
    title={Euclidean and Manhattan distances between Banzhaf indices and voting weights},
    title style = {font=\small},
    xlabel={Threshold (\%)},
    x label style = {font=\small},
    ylabel={Distance of voting power and voting weights},
    y label style = {font=\small},
    xmin=50, xmax=87,
    ymin=0, ymax=0.6,
    legend pos=north west,
    ymajorgrids=true,
    width = 0.98\textwidth,
    height = 0.6\textwidth,
    legend style = {font=\small,at={(0.25,-0.15)},anchor=north west,legend columns=2},
    legend entries = {Manhattan$\qquad$, Euclidean}
]

\addplot[
    color=blue,
    mark=square,
    ]
    coordinates {
(50,0.135280634569506)
(51,0.132498233329874)
(52,0.124364044620391)
(53,0.111477747426217)
(54,0.0947179386854293)
(55,0.0751045872647418)
(56,0.0536615626138057)
(57,0.0330803973515509)
(58,0.0166252962968774)
(59,0.0160594966365823)
(60,0.0343421360175709)
(61,0.0543596236578047)
(62,0.0730940390915708)
(63,0.0904880844668887)
(64,0.106550237505394)
(65,0.121331095597129)
(66,0.134909896542853)
(67,0.147380309367832)
(68,0.158844066407606)
(69,0.169402981573542)
(70,0.179156671268413)
(71,0.192490883793402)
(72,0.207502311717432)
(73,0.223288849127047)
(74,0.241345269153607)
(75,0.259972357497458)
(76,0.279104968710578)
(77,0.298625618708734)
(78,0.318372790288646)
(79,0.342143807082316)
(80,0.367460888322115)
(81,0.393109243118846)
(82,0.418803335427996)
(83,0.444286082771316)
(84,0.470659218106125)
(85,0.49803562559301)
(86,0.525656220152993)
(87,0.557835063654095)

    };
    %\legend{CuSO\(_4\cdot\)5H\(_2\)O}
\addplot[
    color=red,
    mark=square,
]
coordinates{
(50,0.068783968464373)
(51,0.067357801684696)
(52,0.0631898811257396)
(53,0.0565916944669333)
(54,0.0480203821025765)
(55,0.0380090779340954)
(56,0.0271014375140501)
(57,0.0158303932231101)
(58,0.00512890673358288)
(59,0.0077644564431065)
(60,0.0180990117191926)
(61,0.0281125384768308)
(62,0.0375016818773493)
(63,0.0462069572632426)
(64,0.0542278473824966)
(65,0.0615924643202548)
(66,0.0683466430768946)
(67,0.0745455361319342)
(68,0.0802499530870276)
(69,0.0855222016981325)
(70,0.0904248805293897)
(71,0.0950186048081461)
(72,0.0993617361777827)
(73,0.103508689385482)
(74,0.107509537621281)
(75,0.111408031764337)
(76,0.115240776414424)
(77,0.119035357707298)
(78,0.122809811942642)
(79,0.126573045652035)
(80,0.130325703610735)
(81,0.134063289801559)
(82,0.137778857543723)
(83,0.141467062612932)
(84,0.145126621193799)
(85,0.148762992240006)
(86,0.152388136125082)
(87,0.156018338775925)

};

\end{axis}
\end{tikzpicture}
\vspace{0.5cm}
\begin{tikzpicture}
\begin{axis}[
    title={Maximal ratio  between  Banzhaf indices and voting weights},
    title style = {font=\small},
    xlabel={Threshold (\%)},
    x label style = {font=\small},
    ylabel={Distance of voting power and voting weights},
    y label style = {font=\small},
    xmin=50, xmax=87,
    ymin=0, ymax=5,
    legend pos=north west,
    ymajorgrids=true,
    width = 0.98\textwidth,
height = 0.6\textwidth,
legend style = {font=\small,at={(0.35,-0.15)},anchor=north west,legend columns=2},
legend entries = {Maximal ratio}
]  
\addplot[
    color=ForestGreen,
    mark=square,
]
coordinates{
(50,1.41012847907121)
(51,1.40169311068159)
(52,1.37703280024976)
(53,1.33796558648392)
(54,1.28715509989611)
(55,1.22769356637172)
(56,1.1626850371409)
(57,1.09494191396171)
(58,1.02720030065835)
(59,1.04425001599322)
(60,1.11621418383304)
(61,1.19731992780839)
(62,1.28468358972603)
(63,1.37803971198599)
(64,1.4771636140137)
(65,1.58187262435028)
(66,1.69206031138295)
(67,1.8077001165688)
(68,1.92888367244129)
(69,2.05582273843708)
(70,2.18888726407848)
(71,2.32860166978679)
(72,2.47567889993856)
(73,2.63100539513453)
(74,2.79566327721729)
(75,2.97089376294611)
(76,3.1581012248682)
(77,3.35880140587234)
(78,3.5746067812508)
(79,3.80723453016577)
(80,4.05850190199535)
(81,4.33043542369922)
(82,4.62535938431331)
(83,4.94612867533404)
(84,5.29633469631301)
(85,5.68067205398655)
(86,6.10521113968799)
(87,6.5776320481632)

};

\end{axis}
\end{tikzpicture}

\caption{Sum of the distances between the Banzhaf indices and voting weights of the member countries}
\label{Fig2}

\end{figure}

For each quota between 50\% and 87\%, we computed the Euclidean and Manhattan distances, as well as the maximal ratio between the actual voting weights and the Banzhaf indices of all countries. Figure~\ref{Fig2} shows that both distance measures exhibit a unique minimum: the Euclidean distance attains its minimum at a quota of 58\%, while the Manhattan distance reaches its minimum at 59\%. At these threshold levels, the discrepancy between voting weights and voting power is minimised according to absolute distance measures.

At a 58\% threshold, the United States' voting power is 16.93\%, which is very close to its assigned voting weight. While the Euclidean and Manhattan distances emphasise larger absolute deviations, the maximal ratio focuses on relative differences between voting power and voting weight. The maximal ratio also attains its minimum at a threshold of 58\%. At this, the maximum ratio of the Banzhaf power index to the corresponding voting weight is 1.0272, and this maximum is achieved by Japan, which ranks second in voting weights among the member states.

Following \citet{BhattacherjeeSarkar2019}, we compute Pearson's correlation between weights and Banzhaf indices. This approach confirms our results: Figure~\ref{Fig4} shows that the correlation is the highest with a 58\% threshold. 

In the 58-59\% range, the power of the USA is sufficiently low, while the power of the middle and small weighted countries is still smaller than in the case of $q>70\%$.

\citet{Ruff2009} provide an optimal threshold, $q^*=\frac{1}{2}(1+\sqrt{\sum{w_i}})$ for minimizing $\frac{\beta_i}{w_i}$ ratio each player. In the case of IMF, this threshold would be 
60.91\%, which is quite similar to our results using different distance measures.
\begin{figure}[t!]
\centering

\begin{tikzpicture}
\begin{axis}[
    xlabel={Threshold (\%)},
    x label style = {font=\small},
    ylabel={Pearson's correlation},
    xmin=50, xmax=72,
    ymin=0.88, ymax=1,  
    y label style = {font=\small},
    legend pos=north west,
    ymajorgrids=true,
    width = 0.98\textwidth,
height = 0.6\textwidth,
]

\addplot[
    color=blue,
    mark=square,
    ]
    coordinates {
(50,0.979634185910292)
(51,0.980340161968174)
(52,0.982363832709851)
(53,0.985433338077287)
(54,0.989131833118386)
(55,0.992947682707962)
(56,0.996338084115119)
(57,0.998794998499219)
(58,0.99990454008821)
(59,0.999387738819452)
(60,0.997117793876301)
(61,0.993112944309595)
(62,0.987509795556046)
(63,0.980526936029894)
(64,0.972424968522851)
(65,0.963472917247475)
(66,0.953920335390901)
(67,0.943981100052745)
(68,0.933822504901905)
(69,0.923563841063917)
(70,0.913276169594701)
(71,0.902987838023264)
(72,0.892687965802482)
(73,0.882333693496617)
(74,0.871854707480592)
(75,0.861161886058555)
(76,0.850153949709084)
(77,0.838728534330373)
(78,0.826791869162679)
(79,0.814267415792081)
(80,0.801103913744901)
(81,0.787274476979465)
(82,0.772773025336265)
(83,0.757599478832167)
(84,0.741743358806979)
(85,0.725158947573667)
(86,0.707745510083055)
(87,0.689330689477079)

};
 
\end{axis}
\end{tikzpicture}

\caption{Pearson's correlation between the Banzhaf index 
and the voting weights \\ of member countries}
\label{Fig4}

\end{figure}

\begin{figure}[t!]
\centering

\begin{tikzpicture}
\begin{axis}[
    xlabel={Threshold (\%)},
    x label style = {font=\small},
    ylabel={Decisiveness index (\%)},
    y label style = {font=\small},
    xmin=50, xmax=87,
    ymin=0, ymax=60,
    %xtick={0,20,40,60,80,100},
    %ytick={0,20,40,60,80,100,120},
    %legend pos=north west,
    ymajorgrids=true,
    width = 0.98\textwidth,
height = 0.6\textwidth,
]

\addplot[
    color=ForestGreen,
    mark=square,
    ]
    coordinates {
(50,50.0000293106496)
(51,47.0435410441274)
(52,44.0795276425825)
(53,41.1025804695565)
(54,38.1109961884662)
(55,35.1085818820366)
(56,32.1049902932876)
(57,29.1163445528989)
(58,26.1642453832222)
(59,23.2751210320337)
(60,20.4781542888095)
(61,17.8038112677517)
(62,15.2813445565245)
(63,12.9372290470285)
(64,10.7930329573741)
(65,8.86446301320138)
(66,7.160209136148)
(67,5.68198547708014)
(68,4.42453295531314)
(69,3.37659686008887)
(70,2.52183176445239)
(71,1.84030729365739)
(72,1.30980282803313)
(73,0.907312877246399)
(74,0.610207647910006)
(75,0.397315941297574)
(76,0.249616985383312)
(77,0.150729083682553)
(78,0.0870811628298233)
(79,0.0478773119119007)
(80,0.024895639379349)
(81,0.0121545463559117)
(82,0.00552438287551711)
(83,0.00231394656229644)
(84,0.000882447939940116)
(85,0.000301883378467473)
(86,9.09555113993432E-05)
(87,2.35801377302216E-05)

    };
    %\legend{CuSO\(_4\cdot\)5H\(_2\)O}
   
\end{axis}
\end{tikzpicture}

\caption{Power of the Body to Act as a function of quota}
\label{Fig3}

\end{figure}

Figure~\ref{Fig3}  shows that the value of the Power of the Body to Act index decreases sharply with the increase in the quota. At the proposed 58\%  only 26.16\% of the coalitions are winners; a quota of 59\%  would further decrease it to 23.28\%. In the case of the IMF, this steep decline is a well-known phenomenon, \citet{Leech2002b} identified a similar pattern.

\section{Conclusions}\label{Sec5}

The analysis of the IMF's voting system reveals significant disparities between the voting weights and the actual voting power of 191 member countries. Our findings suggest that a decision threshold of 58-59\% would minimise these disparities, aligning voting power more closely with voting weights.

This study underscores the importance of considering decision thresholds in designing fair voting systems. Therefore, future reforms of the IMF's voting system should consider the interplay between voting weights and thresholds to ensure a more equitable distribution of influence among member countries.

From a theoretical perspective, the results suggest that even in large-scale, near-oceanic weighted voting games, the choice of the decision threshold can serve as a powerful second-order design instrument to restore consistency between formal weights and effective power without altering the underlying weight distribution.

\section*{Competing interests}
\addcontentsline{toc}{section}{Competing interests}

The views expressed in this paper are those of the author(s) and do not necessarily reflect the views of the Magyar Nemzeti Bank (MNB).

\section*{Acknowledgements}
\addcontentsline{toc}{section}{Acknowledgements}
We are grateful to \textit{László Csató} for useful feedback, and the anonymous referees for
their valuable comments.

\section*{Funding}
\addcontentsline{toc}{section}{Funding}
The research was supported by the National Research, Development and Innovation Office under Grants PD 153835 and FK 145838.

\bibliographystyle{apalike}
\bibliography{All_references}

\clearpage

\section*{Appendix}
\addcontentsline{toc}{section}{Appendix}
%\section{Voting weights in the IMF}

\setcounter{table}{0}
\renewcommand{\thetable}{A.\arabic{table}}

\begin{footnotesize}
\rowcolors{2}{}{gray!20}
\begin{xltabular}{0.8\textwidth}{Lrrrr}
\caption{Current voting weights and value of Banzhaf indices in the IMF}
\label{Table_A1} \\

\toprule
\multicolumn{1}{L}{Country}& \multicolumn{1}{r}{Weight}& \multicolumn{1}{r}{Rel.\ weight}& \multicolumn{1}{r}{NBI(50\%)}& \multicolumn{1}{r}{NBI(85\%)}  \\ \bottomrule
\endfirsthead

\multicolumn{5}{c}%
{\normalsize{\tablename\ \thetable{} (continued from the previous page)}} \vspace{0.2cm} \\ \toprule
\multicolumn{1}{L}{Country}& \multicolumn{1}{r}{Weight}& \multicolumn{1}{r}{Rel.\ weight}& \multicolumn{1}{r}{NBI(50\%)}& \multicolumn{1}{r}{NBI(85\%)} \\ \bottomrule
\endhead

\toprule \hiderowcolors \multicolumn{5}{r}{{Continued on the next page}} \\ \showrowcolors \bottomrule
\endfoot

\toprule
\multicolumn{5}{l}{Source: \url{https://www.imf.org/en/About/executive-board/members-quotas}}
\endlastfoot
Afghanistan&4690&0.09\%&0.09\%&0.15\%\\
Albania&2845&0.06\%&0.05\%&0.09\%\\
Algeria&21051&0.42\%&0.39\%&0.68\%\\
Andorra&2277&0.05\%&0.04\%&0.07\%\\
Angola&8853&0.18\%&0.16\%&0.29\%\\
Antigua and Barbuda&1652&0.03\%&0.03\%&0.05\%\\
Argentina&33325&0.66\%&0.61\%&1.05\%\\
Armenia&2740&0.05\%&0.05\%&0.09\%\\
Australia&67176&1.33\%&1.23\%&1.92\%\\
Austria&40772&0.81\%&0.75\%&1.27\%\\
Azerbaijan&5369&0.11\%&0.10\%&0.17\%\\
Bahamas&3276&0.06\%&0.06\%&0.11\%\\
Bahrain&5402&0.11\%&0.10\%&0.18\%\\
Bangladesh&12118&0.24\%&0.22\%&0.39\%\\
Barbados&2397&0.05\%&0.04\%&0.08\%\\
Belarus&8267&0.16\%&0.15\%&0.27\%\\
Belgium&65559&1.30\%&1.20\%&1.88\%\\
Belize&1719&0.03\%&0.03\%&0.06\%\\
Benin&2690&0.05\%&0.05\%&0.09\%\\
Bhutan&1656&0.03\%&0.03\%&0.05\%\\
Bolivia&3853&0.08\%&0.07\%&0.13\%\\
Bosnia and Herzegovina&4104&0.08\%&0.08\%&0.13\%\\
Botswana&3424&0.07\%&0.06\%&0.11\%\\
Brazil&111872&2.22\%&2.05\%&2.58\%\\
Brunei Darussalam&4465&0.09\%&0.08\%&0.15\%\\
Bulgaria&10415&0.21\%&0.19\%&0.34\%\\
Burkina Faso&2656&0.05\%&0.05\%&0.09\%\\
Burundi&2992&0.06\%&0.05\%&0.10\%\\
Cabo Verde&1689&0.03\%&0.03\%&0.06\%\\
Cambodia&3202&0.06\%&0.06\%&0.10\%\\
Cameroon&4212&0.08\%&0.08\%&0.14\%\\
Canada&111691&2.22\%&2.04\%&2.58\%\\
Central African Republic&2566&0.05\%&0.05\%&0.08\%\\
Chad&2854&0.06\%&0.05\%&0.09\%\\
Chile&18895&0.37\%&0.35\%&0.61\%\\
China&306281&6.08\%&5.49\%&2.90\%\\
Colombia&21897&0.43\%&0.40\%&0.70\%\\
Comoros&1630&0.03\%&0.03\%&0.05\%\\
Congo, DR&12112&0.24\%&0.22\%&0.39\%\\
Congo, Republic of&3072&0.06\%&0.06\%&0.10\%\\
Costa Rica&5146&0.10\%&0.09\%&0.17\%\\
Côte d'Ivoire&7956&0.16\%&0.15\%&0.26\%\\
Croatia&8626&0.17\%&0.16\%&0.28\%\\
Cyprus&4490&0.09\%&0.08\%&0.15\%\\
Czechia&23254&0.46\%&0.43\%&0.75\%\\
Denmark&35846&0.71\%&0.66\%&1.12\%\\
Djibouti&1770&0.04\%&0.03\%&0.06\%\\
Dominica&1567&0.03\%&0.03\%&0.05\%\\
Dominican Republic&6226&0.12\%&0.11\%&0.20\%\\
Ecuador&8429&0.17\%&0.15\%&0.27\%\\
Egypt&21823&0.43\%&0.40\%&0.70\%\\
El Salvador&4324&0.09\%&0.08\%&0.14\%\\
Equatorial Guinea&3027&0.06\%&0.06\%&0.10\%\\
Eritrea&1611&0.03\%&0.03\%&0.05\%\\
Estonia&3888&0.08\%&0.07\%&0.13\%\\
Eswatini&2237&0.04\%&0.04\%&0.07\%\\
Ethiopia&4459&0.09\%&0.08\%&0.15\%\\
Fiji, Republic of&2436&0.05\%&0.04\%&0.08\%\\
Finland&25558&0.51\%&0.47\%&0.82\%\\
France&203003&4.03\%&3.70\%&2.89\%\\
Gabon&3612&0.07\%&0.07\%&0.12\%\\
Gambia, The&2074&0.04\%&0.04\%&0.07\%\\
Georgia&3556&0.07\%&0.07\%&0.12\%\\
Germany&267796&5.31\%&4.84\%&2.90\%\\
Ghana&8832&0.18\%&0.16\%&0.29\%\\
Greece&25741&0.51\%&0.47\%&0.82\%\\
Grenada&1616&0.03\%&0.03\%&0.05\%\\
Guatemala&5738&0.11\%&0.11\%&0.19\%\\
Guinea&3594&0.07\%&0.07\%&0.12\%\\
Guinea-Bissau&1736&0.03\%&0.03\%&0.06\%\\
Guyana&3270&0.06\%&0.06\%&0.11\%\\
Haiti&3090&0.06\%&0.06\%&0.10\%\\
Honduras&3950&0.08\%&0.07\%&0.13\%\\
Hungary&20852&0.41\%&0.38\%&0.67\%\\
Iceland&4670&0.09\%&0.09\%&0.15\%\\
India&132596&2.63\%&2.43\%&2.73\%\\
Indonesia&47936&0.95\%&0.88\%&1.46\%\\
Iran&37123&0.74\%&0.68\%&1.16\%\\
Iraq&18090&0.36\%&0.33\%&0.58\%\\
Ireland&35951&0.71\%&0.66\%&1.13\%\\
Israel&20661&0.41\%&0.38\%&0.67\%\\
Italy&152152&3.02\%&2.78\%&2.81\%\\
Jamaica&5281&0.10\%&0.10\%&0.17\%\\
Japan&309657&6.14\%&5.54\%&2.90\%\\
Jordan&4883&0.10\%&0.09\%&0.16\%\\
Kazakhstan&13036&0.26\%&0.24\%&0.42\%\\
Kenya&6880&0.14\%&0.13\%&0.22\%\\
Kiribati&1564&0.03\%&0.03\%&0.05\%\\
Korea&87279&1.73\%&1.60\%&2.28\%\\
Kosovo&2278&0.05\%&0.04\%&0.07\%\\
Kuwait&20787&0.41\%&0.38\%&0.67\%\\
Kyrgyz Republic&3228&0.06\%&0.06\%&0.11\%\\
Laos&2510&0.05\%&0.05\%&0.08\%\\
Latvia&4775&0.09\%&0.09\%&0.16\%\\
Lebanon&7787&0.15\%&0.14\%&0.25\%\\
Lesotho&2150&0.04\%&0.04\%&0.07\%\\
Liberia&4036&0.08\%&0.07\%&0.13\%\\
Libya&17184&0.34\%&0.31\%&0.56\%\\
Liechtenstein&2452&0.05\%&0.04\%&0.08\%\\
Lithuania&5868&0.12\%&0.11\%&0.19\%\\
Luxembourg&14670&0.29\%&0.27\%&0.48\%\\
Madagascar&3896&0.08\%&0.07\%&0.13\%\\
Malawi&2840&0.06\%&0.05\%&0.09\%\\
Malaysia&37790&0.75\%&0.69\%&1.18\%\\
Maldives&1664&0.03\%&0.03\%&0.05\%\\
Mali&3318&0.07\%&0.06\%&0.11\%\\
Malta&3135&0.06\%&0.06\%&0.10\%\\
Marshall Islands&1501&0.03\%&0.03\%&0.05\%\\
Mauritania&2740&0.05\%&0.05\%&0.09\%\\
Mauritius&2874&0.06\%&0.05\%&0.09\%\\
Mexico&90579&1.80\%&1.66\%&2.33\%\\
Micronesia&1524&0.03\%&0.03\%&0.05\%\\
Moldova&3177&0.06\%&0.06\%&0.10\%\\
Mongolia&2175&0.04\%&0.04\%&0.07\%\\
Montenegro&2057&0.04\%&0.04\%&0.07\%\\
Morocco&10396&0.21\%&0.19\%&0.34\%\\
Mozambique&3724&0.07\%&0.07\%&0.12\%\\
Myanmar&6620&0.13\%&0.12\%&0.22\%\\
Namibia&3363&0.07\%&0.06\%&0.11\%\\
Nauru&1480&0.03\%&0.03\%&0.05\%\\
Nepal&3021&0.06\%&0.06\%&0.10\%\\
Netherlands&88817&1.76\%&1.63\%&2.31\%\\
New Zealand&13973&0.28\%&0.26\%&0.45\%\\
Nicaragua&4052&0.08\%&0.07\%&0.13\%\\
Niger&2768&0.05\%&0.05\%&0.09\%\\
Nigeria&25997&0.52\%&0.48\%&0.83\%\\
North Macedonia&2855&0.06\%&0.05\%&0.09\%\\
Norway&38999&0.77\%&0.71\%&1.22\%\\
Oman&6896&0.14\%&0.13\%&0.22\%\\
Pakistan&21762&0.43\%&0.40\%&0.70\%\\
Palau&1501&0.03\%&0.03\%&0.05\%\\
Panama&5220&0.10\%&0.10\%&0.17\%\\
Papua New Guinea&4084&0.08\%&0.07\%&0.13\%\\
Paraguay&3466&0.07\%&0.06\%&0.11\%\\
Peru&14797&0.29\%&0.27\%&0.48\%\\
Philippines&21881&0.43\%&0.40\%&0.70\%\\
Poland&42406&0.84\%&0.78\%&1.31\%\\
Portugal&22053&0.44\%&0.40\%&0.71\%\\
Qatar&8803&0.17\%&0.16\%&0.29\%\\
Romania&19566&0.39\%&0.36\%&0.63\%\\
Russian Federation&130489&2.59\%&2.39\%&2.72\%\\
Rwanda&3054&0.06\%&0.06\%&0.10\%\\
Samoa&1614&0.03\%&0.03\%&0.05\%\\
San Marino&1944&0.04\%&0.04\%&0.06\%\\
São Tomé and Príncipe&1600&0.03\%&0.03\%&0.05\%\\
Saudi Arabia&101378&2.01\%&1.86\%&2.47\%\\
Senegal&4688&0.09\%&0.09\%&0.15\%\\
Serbia&8000&0.16\%&0.15\%&0.26\%\\
Seychelles&1681&0.03\%&0.03\%&0.05\%\\
Sierra Leone&3526&0.07\%&0.06\%&0.11\%\\
Singapore&40371&0.80\%&0.74\%&1.25\%\\
Slovak Republic&11462&0.23\%&0.21\%&0.37\%\\
Slovenia&7317&0.15\%&0.13\%&0.24\%\\
Solomon Islands&1660&0.03\%&0.03\%&0.05\%\\
Somalia&3086&0.06\%&0.06\%&0.10\%\\
South Africa&31964&0.63\%&0.59\%&1.01\%\\
South Sudan, Republic of&3912&0.08\%&0.07\%&0.13\%\\
Spain&96807&1.92\%&1.77\%&2.42\%\\
Sri Lanka&7240&0.14\%&0.13\%&0.24\%\\
St.~Kitts and Nevis&1577&0.03\%&0.03\%&0.05\%\\
St.~Lucia&1666&0.03\%&0.03\%&0.05\%\\
St.~Vincent&1569&0.03\%&0.03\%&0.05\%\\
Sudan&7754&0.15\%&0.14\%&0.25\%\\
Suriname&2741&0.05\%&0.05\%&0.09\%\\
Sweden&45752&0.91\%&0.84\%&1.40\%\\
Switzerland&59163&1.17\%&1.08\%&1.74\%\\
Syrian Arab Republic&4388&0.09\%&0.08\%&0.14\%\\
Tajikistan&3192&0.06\%&0.06\%&0.10\%\\
Tanzania&5430&0.11\%&0.10\%&0.18\%\\
Thailand&33571&0.67\%&0.62\%&1.06\%\\
Timor-Leste&1708&0.03\%&0.03\%&0.06\%\\
Togo&2920&0.06\%&0.05\%&0.10\%\\
Tonga&1590&0.03\%&0.03\%&0.05\%\\
Trinidad and Tobago&6150&0.12\%&0.11\%&0.20\%\\
Tunisia&6904&0.14\%&0.13\%&0.22\%\\
Turkmenistan&3838&0.08\%&0.07\%&0.13\%\\
Tuvalu&1477&0.03\%&0.03\%&0.05\%\\
Türkiye&48038&0.95\%&0.88\%&1.46\%\\
Uganda&5062&0.10\%&0.09\%&0.16\%\\
Ukraine&21570&0.43\%&0.40\%&0.69\%\\
United Arab Emirates&24564&0.49\%&0.45\%&0.79\%\\
United Kingdom&203003&4.03\%&3.70\%&2.89\%\\
United States&831394&16.49\%&23.26\%&2.90\%\\
Uruguay&5743&0.11\%&0.11\%&0.19\%\\
Uzbekistan&6964&0.14\%&0.13\%&0.23\%\\
Vanuatu&1690&0.03\%&0.03\%&0.06\%\\
Venezuela&38679&0.77\%&0.71\%&1.21\%\\
Vietnam&12983&0.26\%&0.24\%&0.42\%\\
Yemen, Republic of&6322&0.13\%&0.12\%&0.21\%\\
Zambia&11234&0.22\%&0.21\%&0.37\%\\
Zimbabwe&8520&0.17\%&0.16\%&0.28\%\\

\end{xltabular}
\end{footnotesize}

\end{document}